\documentclass[12pt]{article}

\usepackage{rotating}
\usepackage{wrapfig,epsfig}
\usepackage{psfig}
\usepackage{subfigure}
\usepackage{graphicx}
\usepackage{epic,eepic}
\usepackage{citesort}

\begin{document}

\title{Cluster-Growth in Freely Cooling Granular Media}
\author{S. Luding and H. J. Herrmann\\
Institute for Computer Applications 1, \\
Pfaffenwaldring 27, D-70569 Stuttgart, GERMANY
}
\maketitle

\begin{abstract}
When dissipative particles are left alone, their fluctuation energy decays due 
to collisional interactions, clusters build up and grow with time until the
system size is reached. When the
effective dissipation is strong enough, this may lead to the ``inelastic
collapse'', i.e.~the divergence of the collision frequency of some particles.
The cluster growth is an interesting physical phenomenon, whereas the inelastic
collapse is an intrinsic effect of the inelastic hard sphere (IHS) model used to
study the cluster growth -- involving only a negligible number of particles in
the system. Here, we extend the IHS model by introducing
an elastic contact energy and the related contact duration $t_c$. 
This avoids the inelastic collapse and allows to examine the
long-time behavior of the system. For a quantitative description of
the cluster growth, we propose a burning--like algorithm in continuous space,
that readily identifies all particles that belong to the same cluster. The 
criterion for this is here chosen to be only the particle distance.

With this method we identify three regimes of behavior. First, for short times
a homogeneous cooling state (HCS) exists, where a mean-field theory works nicely,
and the clusters are tiny and grow very slowly. Second, at a certain time 
which depends on the system's properties, cluster growth starts and the 
clusters increase in size and mass until, in the third regime, the system size 
is reached and most of the particles are collected in one huge cluster.
\end{abstract}

\newpage

\section*{Lead paragraph}
Granular media consist of discrete particles and their interaction is
governed by two major concepts:  excluded volume and dissipation.  
Adhesive and frictional forces are neglected in this study 
for the sake of simplicity. Since
the particles are solid, each particle occupies a certain amount of
space, and no other particle may enter this volume.  If another particle
approaches, the pair eventually collides.  During collisions, energy is
lost from those degrees of freedom (linear motion) which
are important for the behavior of the material.  Heat or sound is
radiated and plastic deformation takes place so that energy is
irreversibly gone.

Already such a simple, classical system shows an enormous number of
interesting phenomena like, e.g., shock-waves, size-segregation, surface-waves,
or the clustering in freely cooling systems. The latter will be examined more 
closely here.
Since granular particles dissipate energy, the multi-particle system 
is usually not in equilibrium. This leads to various complex, non-linear
phenomena, as mentioned above. 
In this study, we use a rather simple simulation method for 
inelastic, hard spheres, and examine the time dependent size of ``clusters'', 
i.e.~collections of particles whose nearest-neighbor separation is 
much smaller than the particle size. 
We propose a burning--type algorithm to identify those
particles which belong to one cluster. 
Our new approach is a simple alternative to the analysis involving 
fourier spectra and structure factors or mode coupling-theory.

\section{Introduction}

An essential difference between a classical gas and a granular medium 
is the dissipation of energy. In a classical, conservative system in
equilibrium, energy is conserved. In a dissipative system,
the energy loss is quantified by the restitution coefficient $r$ 
that describes the ratio of the relative velocities of a pair of particles 
after and before the collision so that $r=1$ corresponds to the 
perfectly elastic situation. In dissipative granular systems with $r<1$
and only weak energy input, the phenomenology can be very rich 
\cite{behringer97,herrmann98}.
In this study we focus on the clustering phenomenon.
Starting with a homogeneous density, one observes,
after a short time, the evolution of patterns: Clusters of 
particles form and grow with time \cite{goldhirsch93,mcnamara96}.
This leads to the coexistence of almost empty regions and
areas where the particles are densely packed. This is a  
general scenario in dissipative systems, rather independent of 
the details of the interaction.

The cluster formation may be understood in a qualitative manner:
The homogeneous state contains density and velocity fluctuations.  
Convergent velocity fluctuations can lead to increasing densities in certain 
regions.  As the collision rate increases in these regions, so does the energy
dissipation rate. If the energy dissipation rate is great enough,
the pressure cannot reverse the convergent flow and, if this process is
not terminated by sufficiently strong perturbations, it is
self-stabilizing, causes clusters and may eventually lead to the
``inelastic collapse''. Note that inelastic collapse is not another
expression for clustering, rather it describes the divergence of 
the collision frequency in the system, possibly involving a few particles
only.

The size of the clusters, and thus one of the typical lengths of
the system, grows until the system size is reached
\cite{goldhirsch93,mcnamara96,luding98f}. 
The beginning of cluster growth can also be explained by means of
a hydrodynamic stability analysis \cite{mcnamara96,brey98}.
Clusters build up only if the effective density or dissipation
is large enough. In the other case, the fluctuations of the kinetic
energy can act as perturbations and destroy evolving clusters so that
the system stays homogeneous. In intermediate situations shear modes
are observed, for example see \cite{mcnamara96}.

The clustering instability and the inelastic collapse were carefully
examined in 1D
\cite{bernu90,mcnamara92,mcnamara93,luding94,grossman96b,kudrolli97,kudrolli97b}, and
in 2D
\cite{sibuya90,goldhirsch93,mcnamara94,trizac95,mcnamara96,spahn97,deltour97,orza97,luding98f}; 
systematic three dimensional studies were not performed to our knowledge.
Dissipation reduces the fluctuation velocities and thus reduces the free volume. 
A decreasing free volume can lead to disorder--order transitions 
\cite{olafsen98} and the inelastic collapse \cite{mcnamara96}. 
The inelastic collapse is due to the
perfectly rigid interaction potential of the particles in the
frequently used inelastic hard-sphere (IHS) model. 
Those potentials allow instantaneous contacts and thus a 
diverging number of collisions within small time-intervals. 

In theoretical approaches, simplifying assumptions are made to keep the
system trac\-table, these are, e.g., constant density, weak dissipation,
and vanishing gradients of the field variables 
\cite{haff83,savage79,jenkins79,brito98}.
A phenomenon like cluster growth cannot be described easily
in the case of strong dissipation, where it is usually observed.
Even if the global density is small and initially homogeneous, the 
density can get large locally and large gradients at the boundary 
between the clusters and the ``vacuum'' exist. 
For low densities and weak dissipation, the predictions of kinetic
theory and stability analysis are true \cite{mcnamara96} and in the case of a 
homogeneous density, a perfect agreement between numerical simulations and 
kinetic theory was obtained for inelastic, rough, spherical particles
in 2D and 3D also for high densities \cite{luding98e}. 

In order to model clustering, the inelastic
hard sphere (IHS) model is frequently used, since
the event driven nature of the modelling algorithm allows simulations
over many orders of magnitude in time \cite{luding98e,luding98f}. 
The only drawback is the
occurence of the inelastic collapse due to the rigid nature of contacts.
In the framework of the IHS, the inelastic collapse is related to a 
divergence of the collision frequency and a vanishing particle separation 
and relative velocity. Different possibilities to avoid the unphysical 
inelastic collapse were proposed, and are reviewed in Ref.\ \cite{luding98f}. 
One of them introduces a virtual time of contact (TC), as also used in the 
following. The so-called TC model was examined in detail and seems to influence 
only a small fraction of the particles in the system \cite{luding98f}. 
The idea is to introduce an 
elastic energy which cannot be dissipated. When two particles are in contact,
energy is stored in the deformed contact volume. This contact energy is
not dissipated, but eventually converted into kinetic energy when the
particles separate.  In the TC model, particles that collide too
frequently, i.e. more than once in a time-interval $t_c$, are assumed to
be in contact, and their kinetic energy is not dissipated by further
collisions.  The time $t_c$ can be identified as the contact duration.
The contact energy cannot be dissipated so that the collision frequency 
cannot diverge further, which implies that also the relative velocity and the 
separation cannot decrease further.

Using the TC model, vibrated \cite{luding96e,luding97c} and freely 
cooling \cite{luding98f} systems were examined in detail. For sufficiently
weak dissipation, the homogeneous cooling state (HCS) is observed. 
In the HCS only a negligible number of particles collide twice within 
a short time $t_c$.  If dissipation becomes stronger, clusters build up and
the collision frequency inside the clusters increases. In this situation, the 
TC model is active, i.e.~dissipation is switched off during a considerable
number of collisions, and thus limits the steady increase of the collision 
frequency in the centers of the clusters.
However, even when the collision frequency is strongly influenced by
the TC model, the global cooling, i.e.~the energy of the system, is
almost unaffected.
Furthermore, one observes that the TC model is able to avoid the
inelastic collapse in the case of very strong dissipation, i.e.~very
small restitution coefficients $r$, and also in the case of large
systems, where clusters can grow correspondingly \cite{luding98f}.

In the following, we use the TC model extension of the IHS to 
investigate the freely cooling granular system. The homogeneous
cooling state is discussed in section\ \ref{sec:HCS}, and the
influence of the parameters $r$ and $t_c$ is examined using small
systems in section\ \ref{sec:FCS}. Cluster growth in large systems is
presented in section\ \ref{sec:CGrow}, and its quantitative analysis 
with a ``burning''-type algorithm is introduced in section\ \ref{sec:Burning}.

\section{The homogeneous cooling state}
\label{sec:HCS}

In order to examine a system with homogeneous density, it
is convenient to use periodic boundary conditions \cite{allen87}. 
In a periodic, freely cooling system, given an isotropic initial
condition, each point is a-priori identical to every other point, 
besides fluctuations in the initial configuration. In such a system
without walls and without externally applied gradients in, say, 
temperature, the system has the chance to stay homogeneous; 
the presence of walls or other external forces can cause 
inhomogeneities which are thus due to the boundaries, but not
intrinsic to the material.  If the system is prepared
with a homogeneous density and, as usual, relaxed for 
a certain time without dissipation, $r=1$, so that a 
Maxwell-Boltzmann velocity distribution is achieved, 
one has a starting configuration that allows examination of both the
homogeneous cooling state (HCS) and the growth of clusters.  The HCS
occurs as long as dissipation is small, and fluctuations remain strong enough 
to disrupt the formation of clusters.

The dimensionless kinetic energy $K=E(t)/E(0)$ of a freely cooling
system (where $E(t)$ is the kinetic translational energy), is well 
predicted by the analytical solution for the HCS of smooth particles 
(see Ref.\ \cite{luding98d} and references therein):
\begin{equation}
K = \left ( {1+\frac{1-r^2}{4} ~ \tau} \right )^{-2}~.
\label{eq:HCS}
\end{equation}
In Eq.\ (\ref{eq:HCS}) $\tau=t_E^{-1}(0)\,t$ is the time rescaled 
with the Enskog collision rate
\begin{equation}
t_E^{-1}(t) = 2 d \frac{N}{V} \sqrt{\pi \frac{E(t)}{M}} \, g(2a)
\end{equation}
at time $t=0$, with the particle diameter $d$, the particle number $N$, the system
volume $V$, the total particle mass $M=N m$, and the particle-particle correlation 
function at contact $g(2a)$ \cite{luding98f}. 
The decay of energy in the rescaled time frame depends only on $r$, and 
all dependencies on quantities like system size and density are hidden
in $\tau$.

\section{Effect of $r$ and $t_c$}
\label{sec:FCS}

In the following, systems of length $L = l/d$ with $N$ particles
of diameter $d$ are examined with event driven (ED) simulations
\cite{luding98e,luding98f} in two dimensions (2D).
The volume fraction is defined as $\varrho = N \pi (d/2 l)^2 = 
(\pi/4) N/V$. Initially the particles are arranged on a square
lattice with random velocities drawn from an interval with constant
probability for each coordinate. The mean total velocity, i.e.~the random
momentum due to the fluctuations, is disregarded in order to have a system 
with its center of mass at rest. 
The system is evolved for some time, until the arbitrary initial condition 
is forgotten, i.e.~the density is homogeneous, and the velocity distribution 
is a Gaussian in each coordinate.

\begin{figure}[tbp]
{(a)} ~\hspace{-1cm}~ \epsfig{file=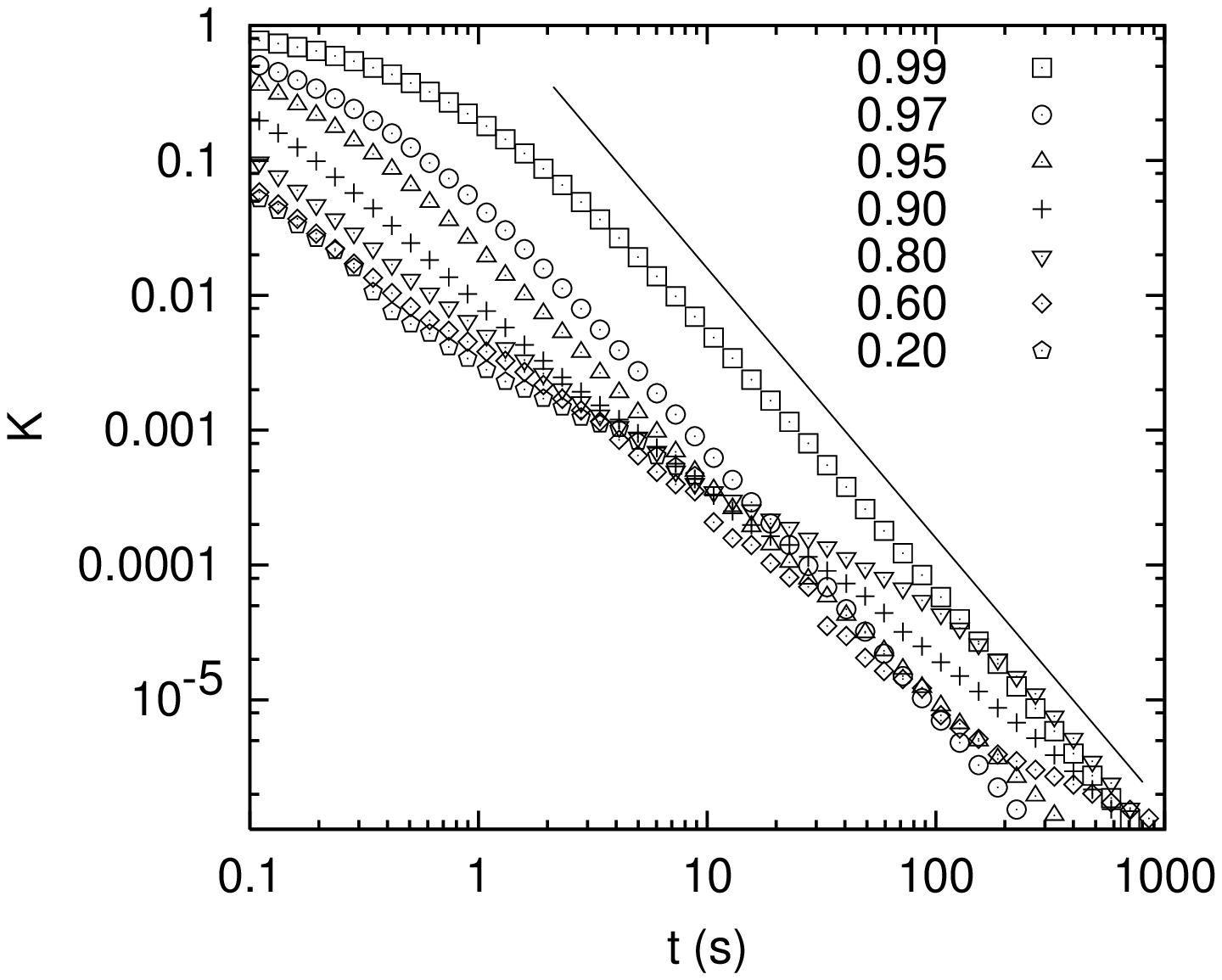,height=7.5cm,angle=-90} \hfill
{(b)} ~\hspace{-1cm}~ \epsfig{file=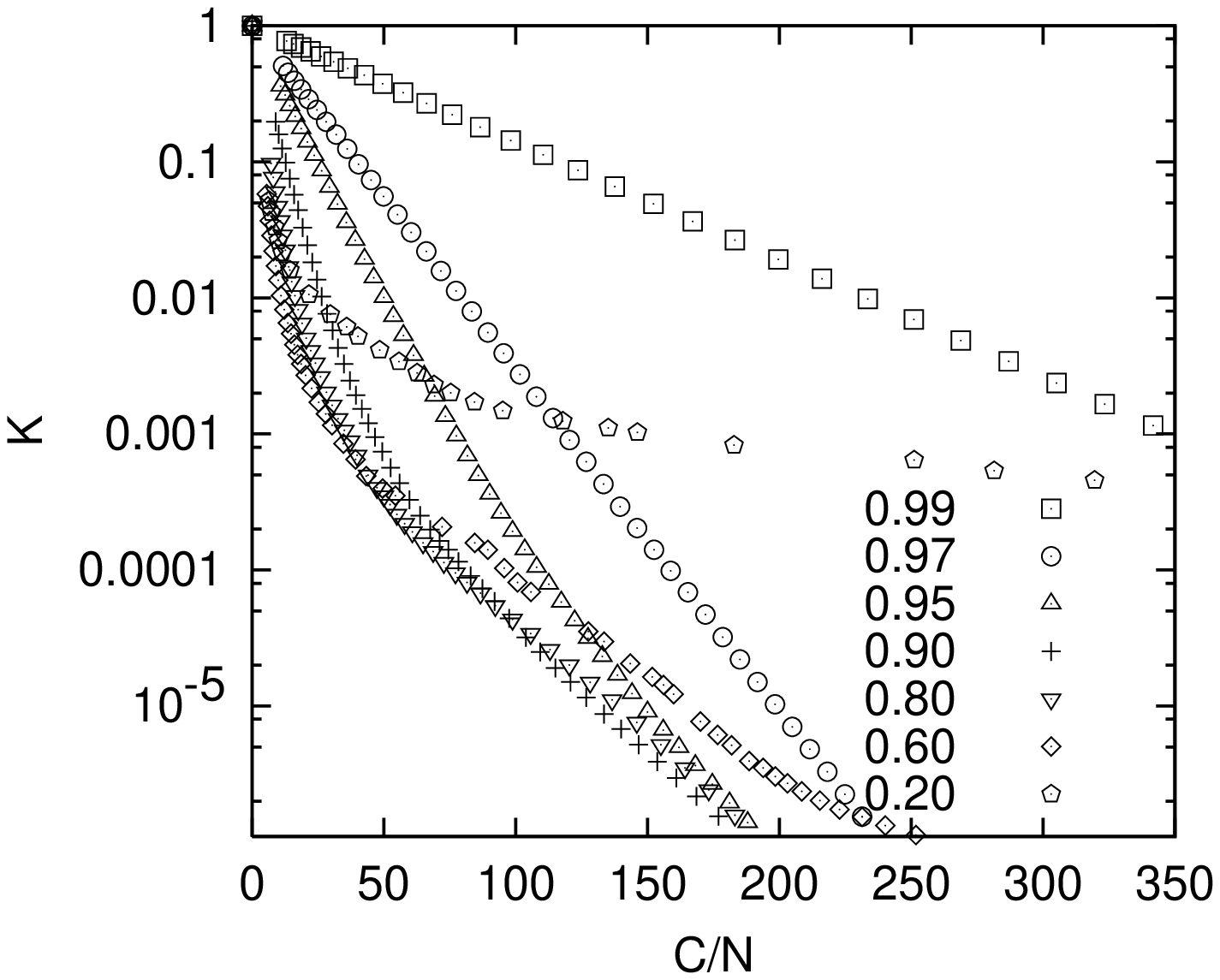,height=7.5cm,angle=-90} \\
{(c)} ~\hspace{-1cm}~ \epsfig{file=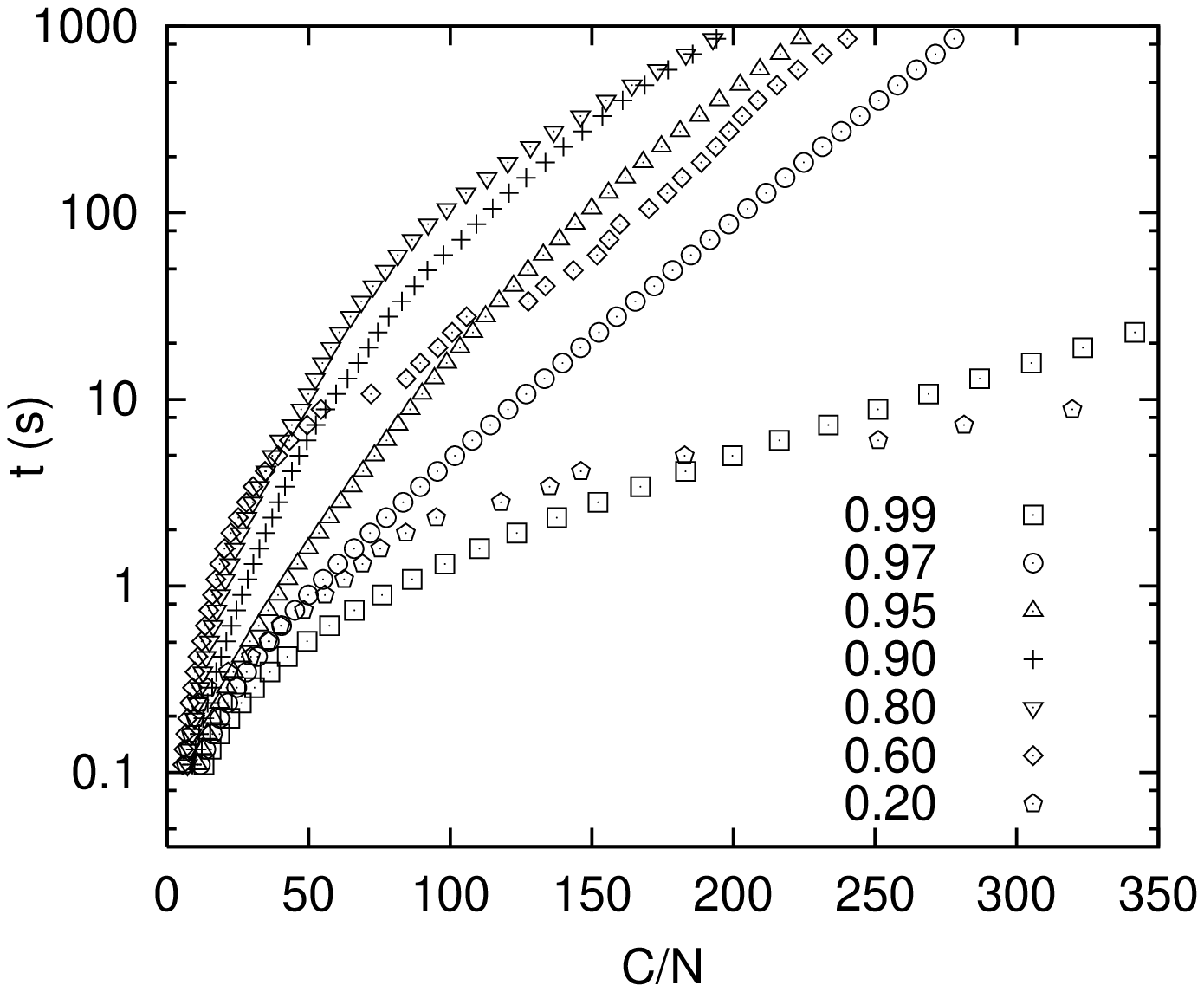,height=7.5cm,angle=-90} \hfill
{(d)} ~\hspace{-1cm}~ \epsfig{file=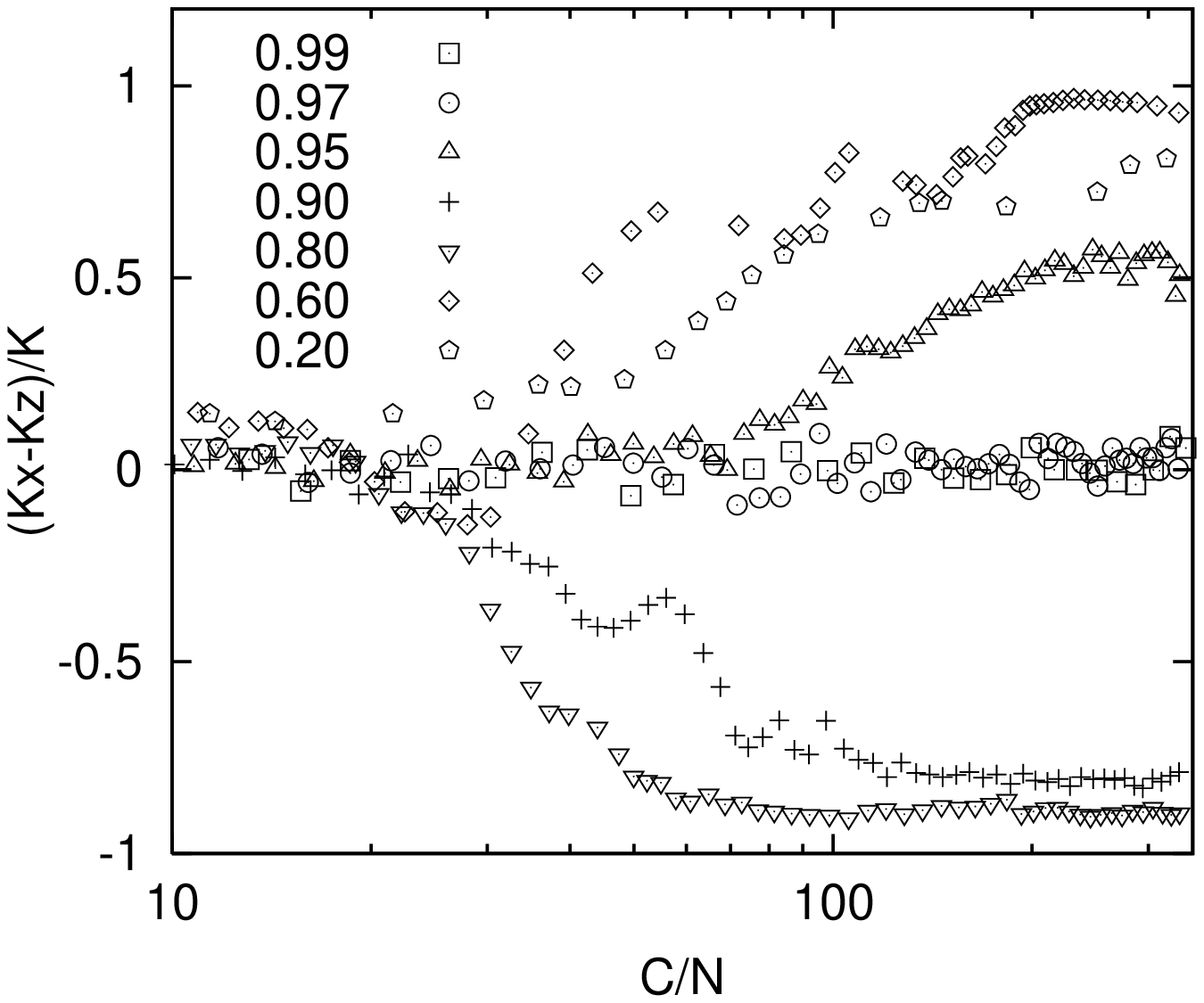,height=7.5cm,angle=-90}
\caption{
(a) Double-logarithmic plot of the dimensionless energy $K$ as function of 
time $t$ for different restitution coefficients $r$ as given in the inset. 
For the other simulation parameters see the text. The solid line indicates
the $K \propto t^{-2}$ behavior in the HCS.
(b) Log-linear plot of $K$ against $C/N$ from the same simulations as in (a).
(c) Log-lin plot of $t$ as function of $C/N$.
(d) Linear-logarithmic plot of $(K_x-K_z)/K$ as function of $C/N$.
}
\label{fig:ene01}
\end{figure}

From those initial configurations, the simulation is
started with $r < 1$ at time $t=0$. In 2D the dimensionless kinetic 
energy $K = E(t)/E(0) = K_x+K_z$ has two contributions, one from
each direction $x$ and $z$. If the system is isotropic $K_x \approx K_z$,
and on the other hand, if shear modes or clusters are present, one
of the components usually dominates. The first set of simulations concerns
a small system with $N=784$, $L=50$, $t_c=10^{-5}$\,s, and 
$\varrho \approx 0.25$. 
Simulations are performed for different $r = 0.99$, 0.97, 0.95, 0.9, 0.8, 
0.6, and 0.2 until every particle carried out $C/N = 1000$ collisions 
on average. The initial collision rate
for these simulations was $t_E^{-1}(0)=251.2$\,s$^{-1}$. 
In Fig.\ \ref{fig:ene01}(a), $K$ is plotted against
time $t$, while in Fig.\ \ref{fig:ene01}(b) it is plotted against the
accumulated mean number of collisions per particle $C/N$. 
Only the simulations with $r = 0.99$ (squares)
and $r = 0.97$ (circles) are described well by the theoretical
solution of the HCS in Eq.\ (\ref{eq:HCS}). For smaller
values of $r$, i.e.~stronger dissipation, the simulations cool more slowly
than predicted by the corresponding analytical solution. This is due to 
the build-up of clusters, where many particles move together while their
relative velocity is rather small. For large $r$ 
the energy behaves as 
\begin{equation}
K = \exp \left [- \frac{1-r^2}{2} \frac{C}{N}  \right ] ~,
\label{eq:KC}
\end{equation}
as can be derived using Eq.\ (\ref{eq:HCS}) for the integration of the
collision rate over time $t$ or the normalized fluctuation
velocity over $\tau$: 
\begin{equation}
\frac{C}{N} = \int_0^t t_E^{-1}(t) dt = \int_0^t \sqrt{K} d\tau
            = \frac{4}{1-r^2} \ln \sqrt{K^{-1}} ~.
\end{equation}
If we were to plot Eq.\ (\ref{eq:KC}) then it would appear as a straight line 
in Fig.\ \ref{fig:ene01}(b). Such a behavior is observed for $r \approx 1$ or 
for small $C/N$, but already for $r \le 0.95$ the simulations deviate 
from the theoretical prediction. 

Note that the system dissipates energy
slower for small $r$ and rather large times $t$. 
This non-intuitive result can be explained as follows. 
For small times, the simulations follow the theoretical prediction, 
the smaller $r$, the stronger is the decay of energy. 
However, for smaller $r$ the deviation from Eq.\ (\ref{eq:HCS}) starts 
earlier, since clusters can build up and grow faster for smaller $r$.
As soon as the clusters have reached a size, so that an incoming
particle (or another cluster) is practically absorbed 
($r_{\rm eff} \approx 0$) the behavior will no longer directly 
depend on $r$. Naturally, the size of a cluster,
necessary for an effective restitution of $r_{\rm eff} \approx 0$, is smaller
for stronger dissipation and thus smaller $r$. The fluctuations from 
one simulation to the other are quite large in Fig.\ \ref{fig:ene01}(a), due 
to the small systems. Other simulations with larger systems show that the 
variations become less important, and that the system's energy is in fact 
almost independent of the value of $r$ for intermediate to large times 
\cite{luding98f}. 

In Fig.\ \ref{fig:ene01}(c) the real-time $t$ is plotted against the 
system inherent accumulated number of collisions per particle $C/N$.
The latter grows faster for weaker dissipation, since in this situation
the relative energy remains longer in the system so that the collision 
rate decays more slowly. 
Following the estimate of a critical restitution coefficient,
see Eq.\ (9) in Ref.\ \cite{mcnamara96}, the inelastic
collapse can be expected if $r$ is smaller than
\begin{equation}
r_c(N,\varrho) \approx \tan ^2 \left [ \frac{\pi}{4}
            \left ( 1-\frac{1}{\lambda_{\rm opt}} \right ) \right ]~,
\label{eq:rc}
\end{equation}
with the non-dimensional optical depth
$\lambda_{\rm opt} = \sqrt{\pi N \varrho}/2$.
In the opposite case when $r > r_c(N,\varrho)$ the system remains in the
homogeneous cooling state, whereas in the intermediate regime, density
variations occur but do not in general lead to a divergence of the collision
frequency.

In Figs.\ \ref{fig:ene01}(b-d), horizontal jumps from one data-point 
to the next indicate clusters and cluster-collisions, with many collisions 
occuring within a short time, and thus increasing $C/N$ discontinuously. 
The energy $K$ is not affected in the same way, since only a few particles 
carry out the large number of collisions.  Note that the TC model is 
activated when these few particles collide more than once in the 
time-interval $t_c$.
In Fig.\ \ref{fig:ene01}(d), the fraction of
energy in the two coordinate directions is compared.
The difference $(K_x-K_z)/K$ lies always between $-1$ and $+1$.
If $(K_x-K_z)/K > 0$ the major part of the energy is contained
in the horizontal degree of freedom, otherwise, if $(K_x-K_z)/K < 0$,
the larger part contributes to vertical motion. Thus $(K_x-K_z)/K$
is a measure for the inhomogeneity of the system concerning convective
motion. It is evident,
that a deviation from the theoretical prediction for the HCS 
is correlated to non-zero values of $(K_x-K_z)/K$. The lower $r$
is, the earlier the deviations begin.

\begin{figure}[tbp]
{(a)} ~\hspace{-1cm}~ \epsfig{file=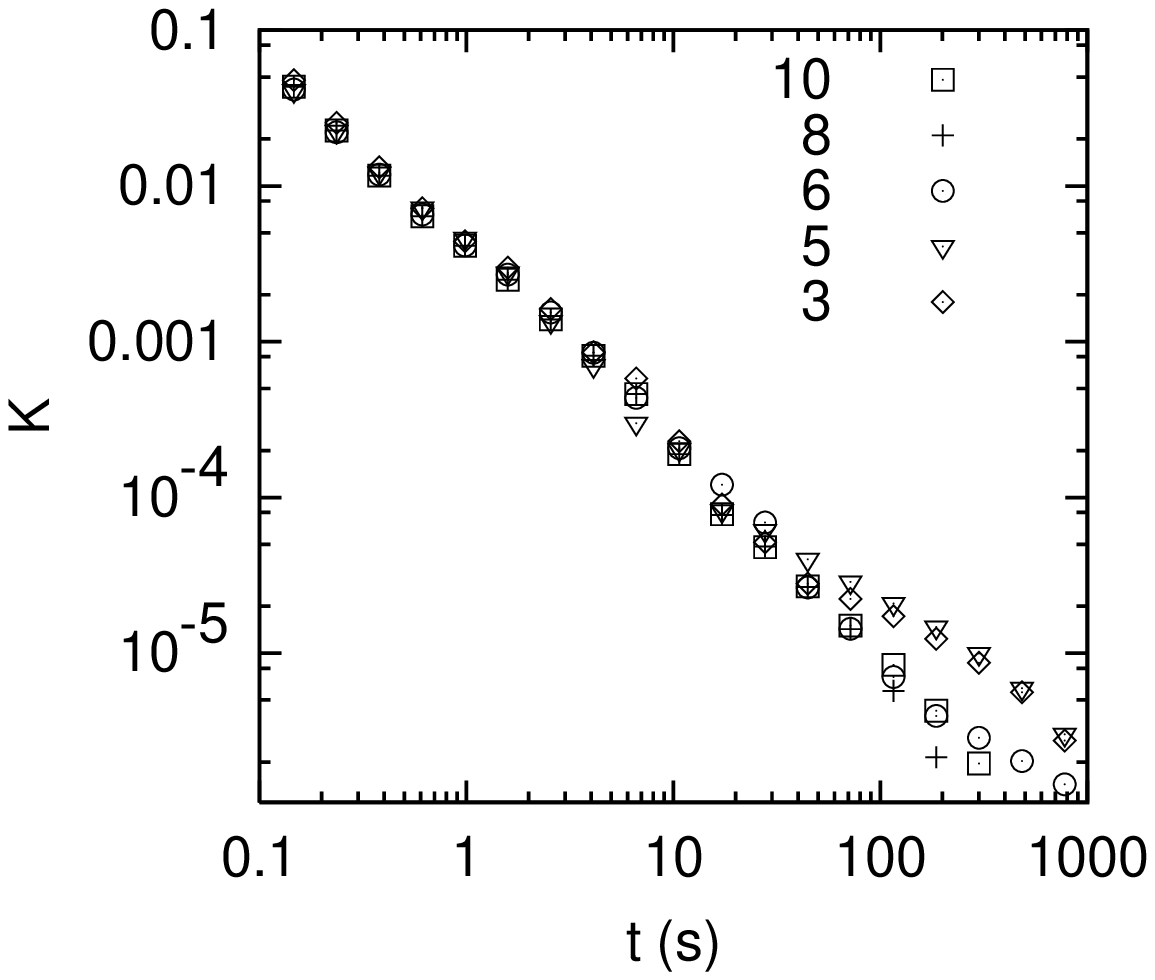,height=7.5cm,angle=-90} \hfill
{(b)} ~\hspace{-1cm}~ \epsfig{file=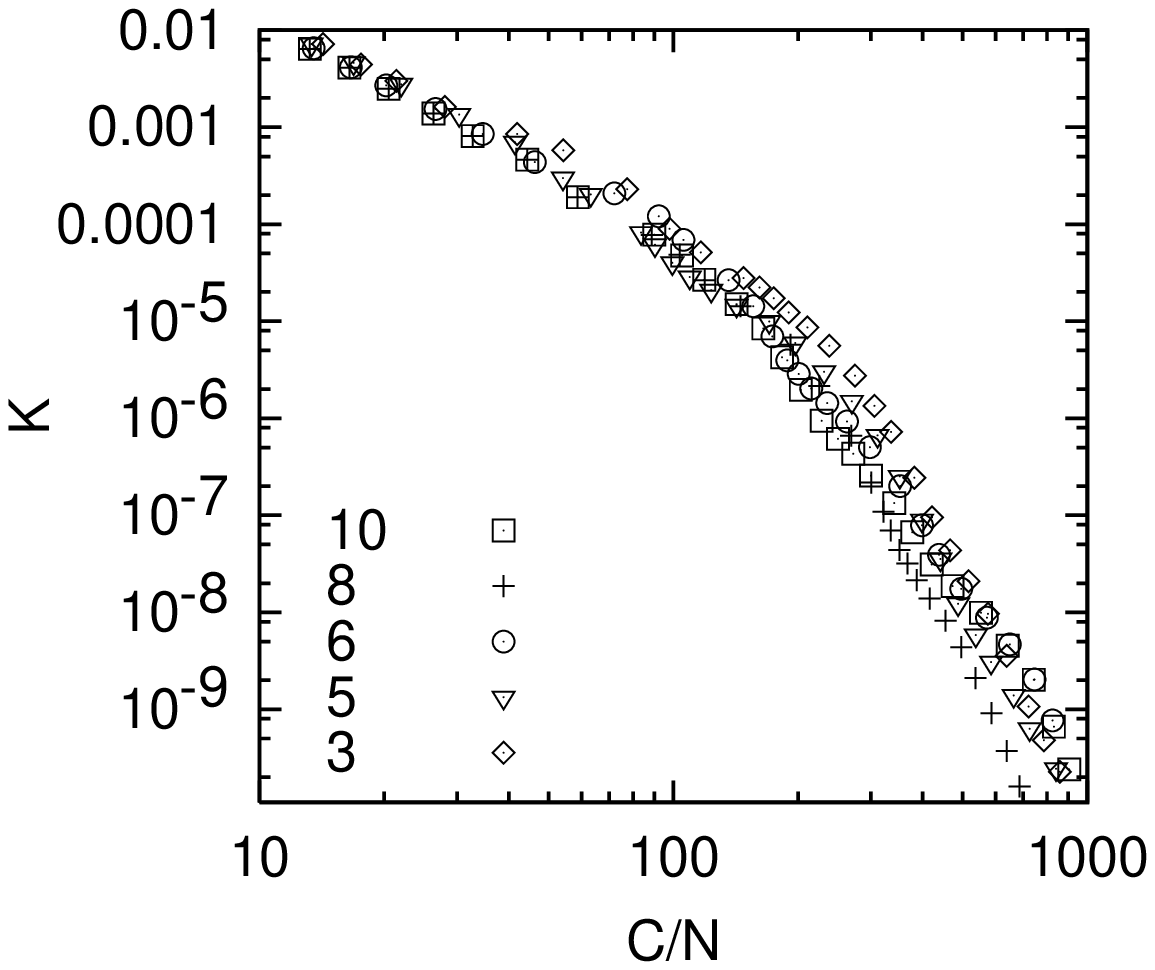,height=7.5cm,angle=-90} \\
{(c)} ~\hspace{-1cm}~ \epsfig{file=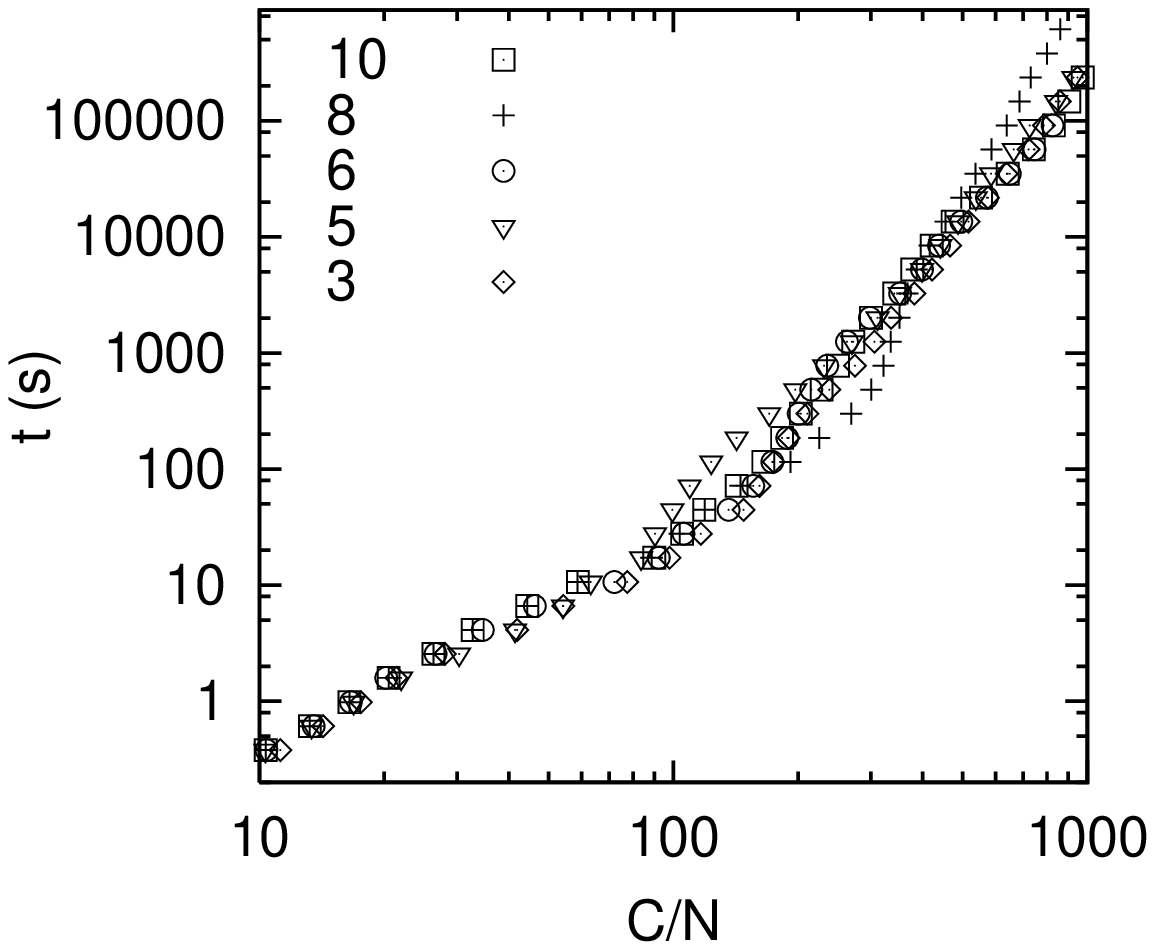,height=7.5cm,angle=-90} \hfill
{(d)} ~\hspace{-1cm}~ \epsfig{file=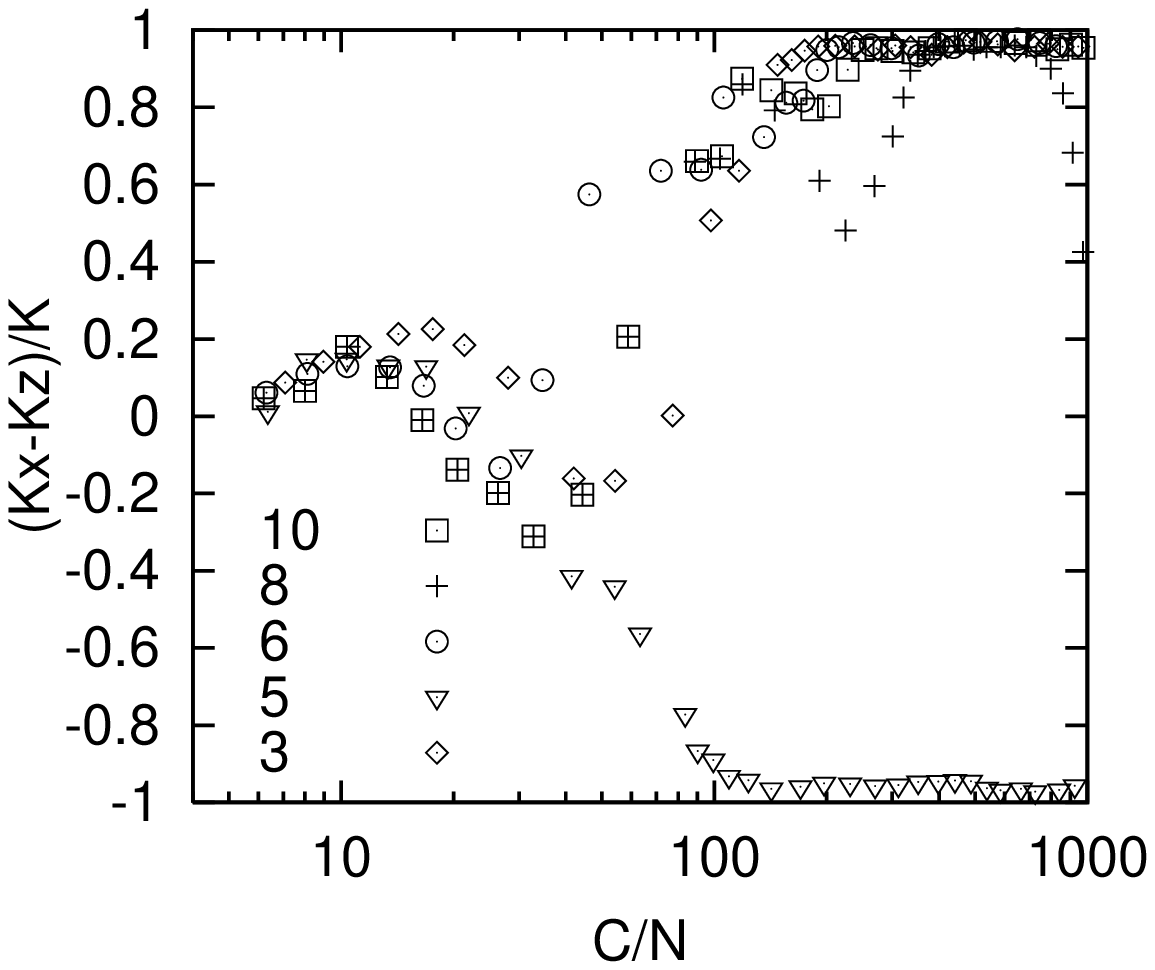,height=7.5cm,angle=-90} 
\caption{
(a) Log-log plot of $K$ against $t$ for different contact durations 
$t_c$, with $-\log_{10} t_c$ given in the inset. 
For the other simulation parameters see the text.
(b) Log-log plot of $K$ against $C/N$ from the same simulations.
(c) Log-log plot of $t$ as function of $C/N$.
(d) Lin-log plot of $(K_x-K_z)/K$ as function of $C/N$.
}
\label{fig:ene02}
\end{figure}

For a better understanding in how far the TC model changes the
system behavior, the simulation with $r=0.6$ is examined for
different $t_c = 10^{-10}$\,s, $10^{-8}$\,s, $10^{-6}$\,s, $10^{-5}$\,s
and $10^{-3}$\,s. In Fig.\ \ref{fig:ene02} the combinations of
$K$, $t$, $C/N$, and $(K_x-K_z)/K$ are displayed as in Fig.\ \ref{fig:ene01}.
From Figs.\ \ref{fig:ene02}(a-c), it follows that the system behavior
does not depend on $t_c$ for small $C/N$ and short times.
Also, the global kinetic energy is almost independent of $t_c$ for large
times, as long as $t_c < 10^{-5}$\,s.
For $C/N > 40$, a shear mode builds up, see Fig.\ \ref{fig:ene02}(d),
and the orientation, i.e.~the value of $(K_x-K_z)/K$, depends strongly
on $t_c$, since tiny changes due to a modiefied $t_c$ may lead to
rather random situations from one realization to the next, even when
identical initial conditions are used.

As soon as strong fluctuations in density exist, a comparatively
large number of particles is affected by the TC model and the
details of the behavior of the system depend on $t_c$.
One sees the tendency in Fig.\ \ref{fig:ene02}(a) that larger values of
$t_c$ lead to a weaker dissipation during time. On the other hand,
especially the simulation with $t_c = 10^{-8}$\,s deviates strongly
from the other ones, see Fig.\ \ref{fig:ene02}(b) and (c). 
For a more detailed discussion of the effect of $t_c$ on the system 
behavior see Ref.\ \cite{luding98f}.

\section{Cluster Growth}
\label{sec:CGrow}

For $r < 0.97$ the system becomes inhomogenous quite rapidly.
Clusters, and thus also dilute regions, build up and have the tendency 
to grow. Since the system is finite, their extension will reach system 
size at a finite time. Thus we distinguish between three phases
of development of the system: (i) the initially (almost) homogeneous state,
(ii) the cluster growth regime, and (iii) the system size dependent final 
stage where the clusters have reached system size.

\begin{figure}[htbp]
{(a)} ~\hspace{-1cm}~ \epsfig{file=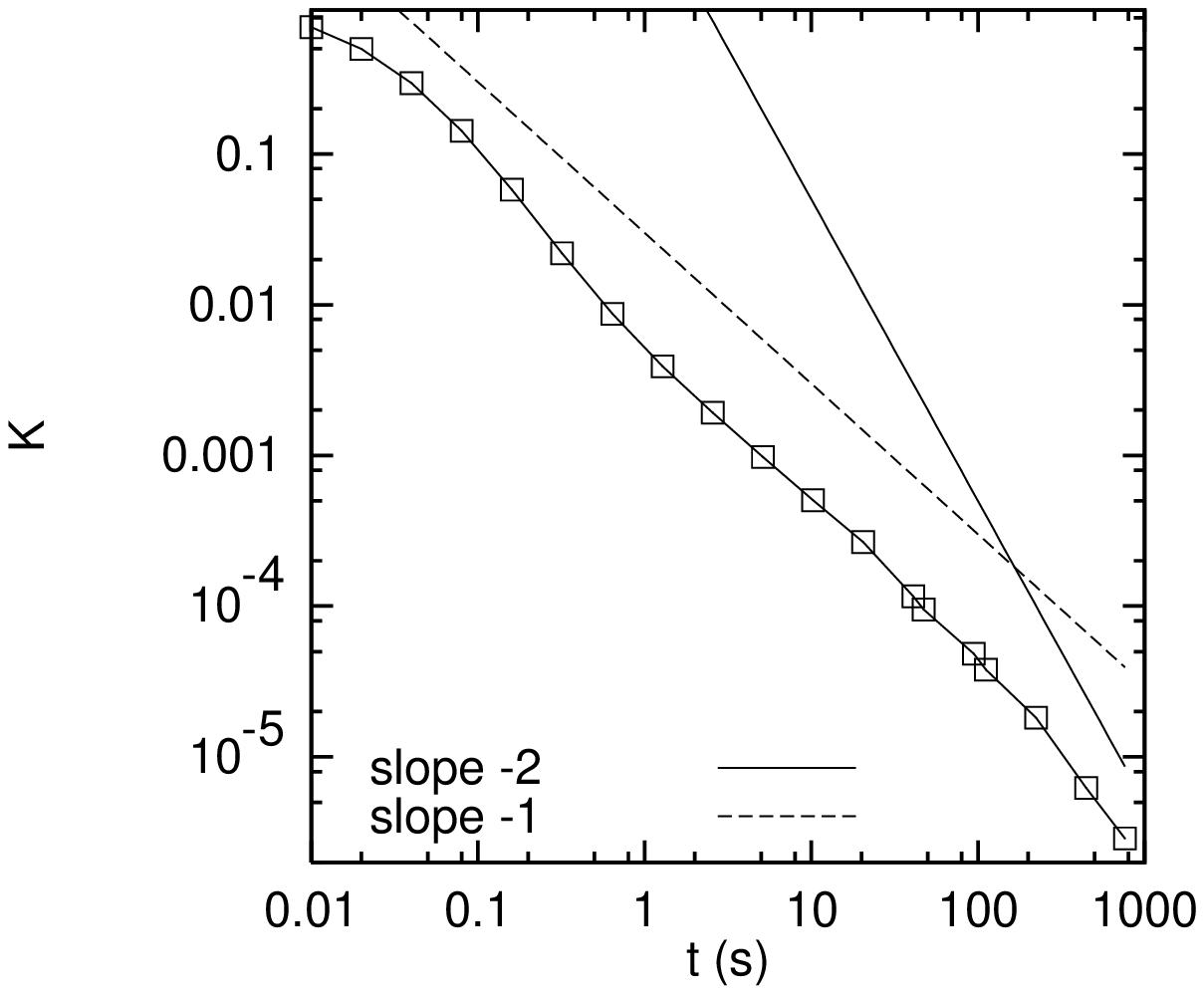,height=7.5cm,angle=-90} \hfill
{(b)} ~\hspace{-1cm}~ \epsfig{file=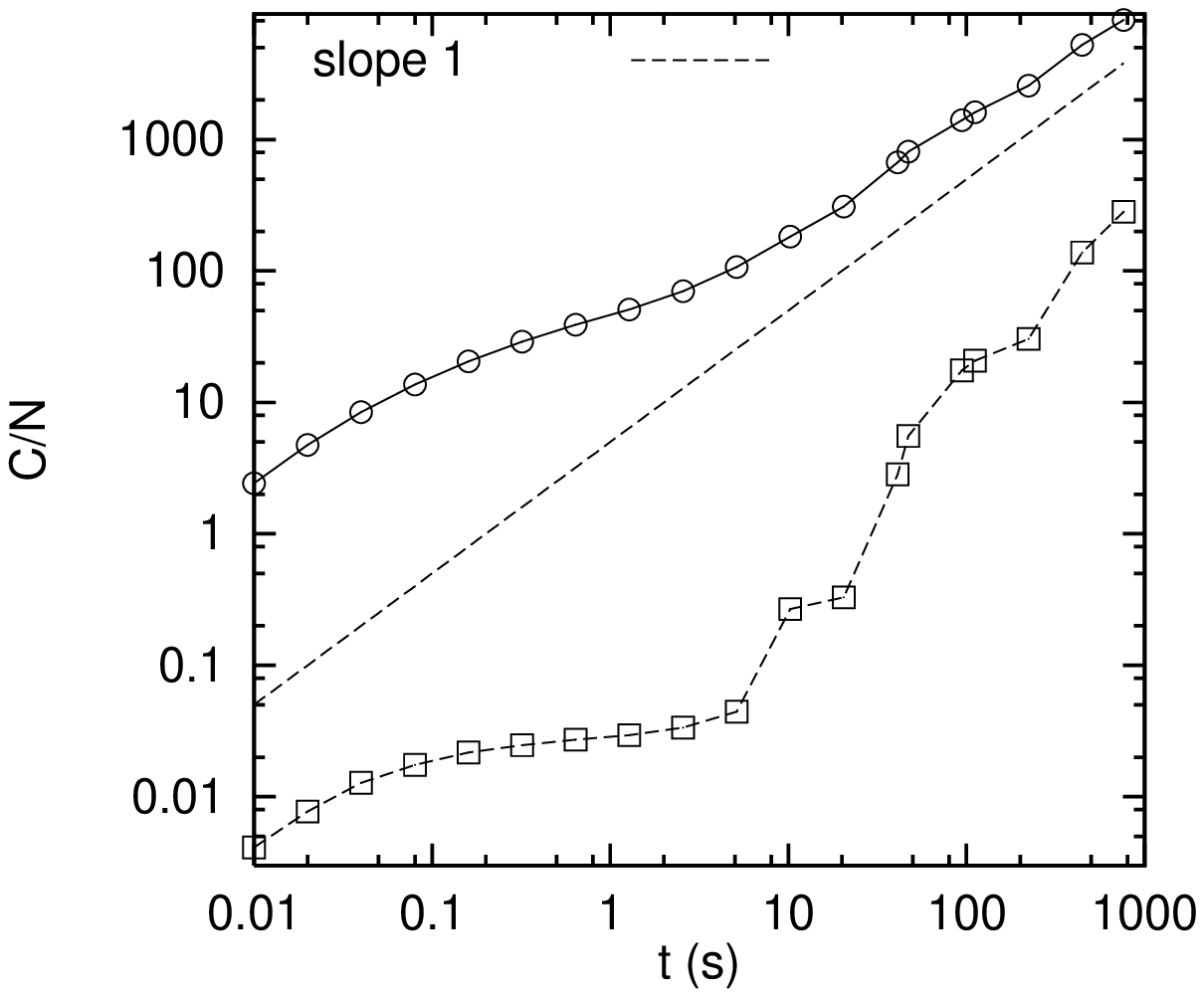,height=7.5cm,angle=-90} 
\caption{
(a) Log-log plot of $K$ against $t$ for a large simulation, for details
see the text.
(b) Log-log plot of $C/N$ against $t$ for the same simulation as in (a). 
The circles correspond to $C/N$ and the squares give the accumulated number 
of collisions per particle when the TC model was active, i.e.~for which the
involved particles collided more than once within time $t_c$.
}
\label{fig:cluster_K}
\end{figure}

In order to examine the cluster growth regime we turn to a much
larger system with $N=79524$, $L=500$, $\varrho = 0.25$, $r=0.8$, and
$t_c = 10^{-5}$\,s. In Fig.\ \ref{fig:cluster_K} the energy $K$
and the total number of collisions $C/N$ are displayed as functions
of time $t$.

\begin{figure}[htbp]
{$t=0.640$ s, $C/N=39$} \hfill {$t=2.56$ s, $C/N=70$} \\
\epsfig{file=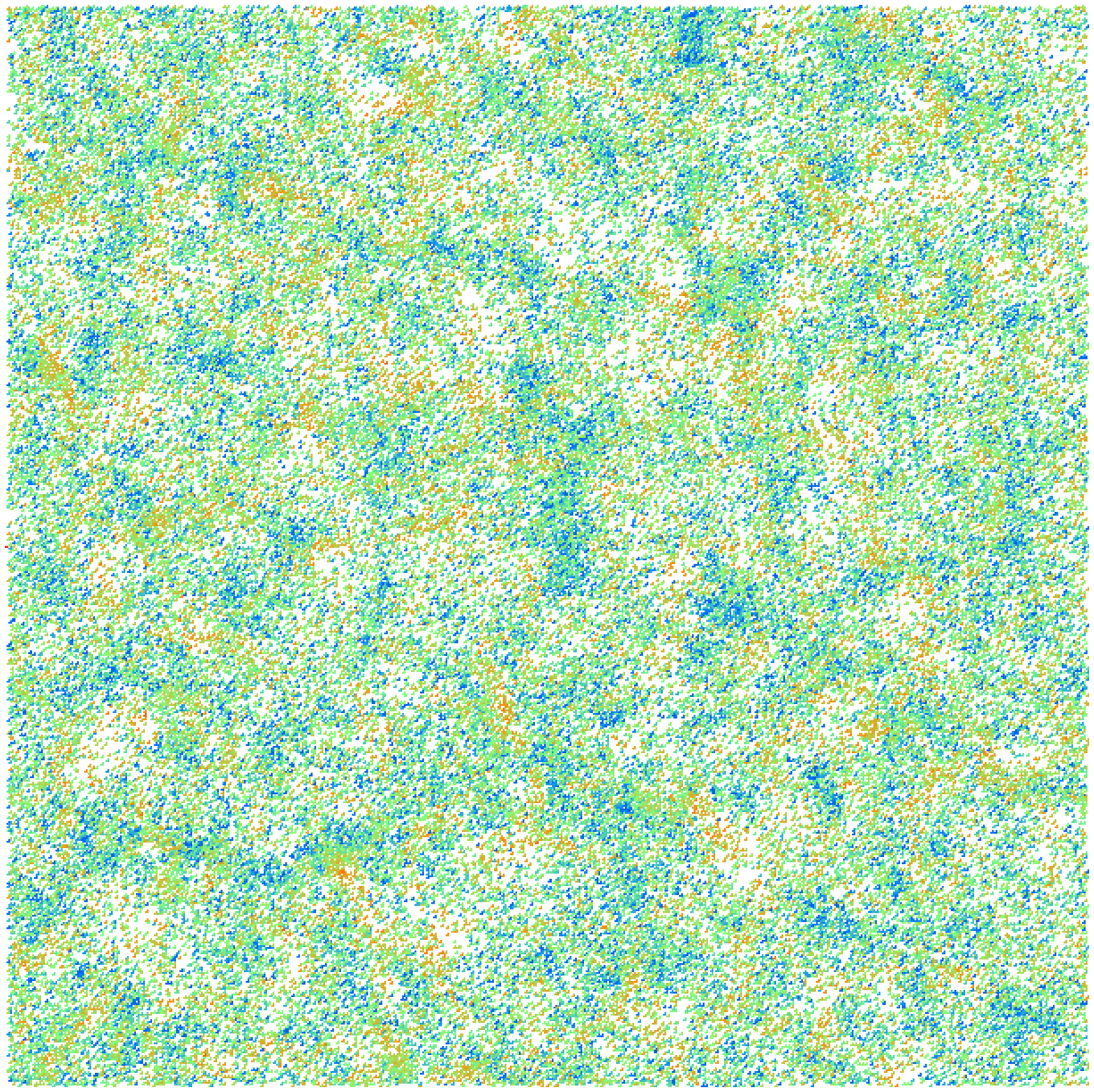,height=7.7cm} \hfill
\epsfig{file=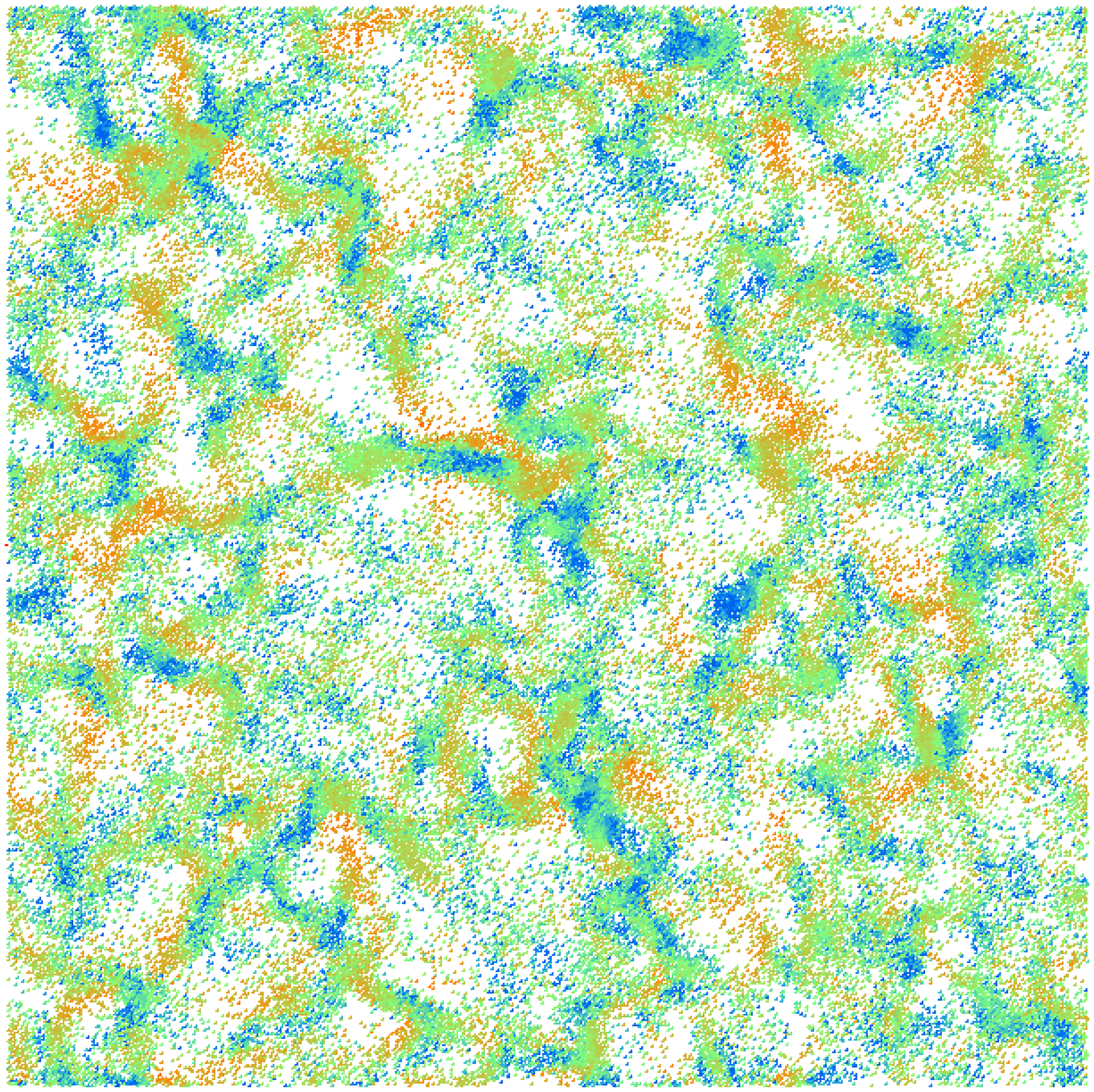,height=7.7cm}
{$t=40.96$ s, $C/N=670$} \hfill {$t=446.6$ s, $C/N=5258$} \\
\epsfig{file=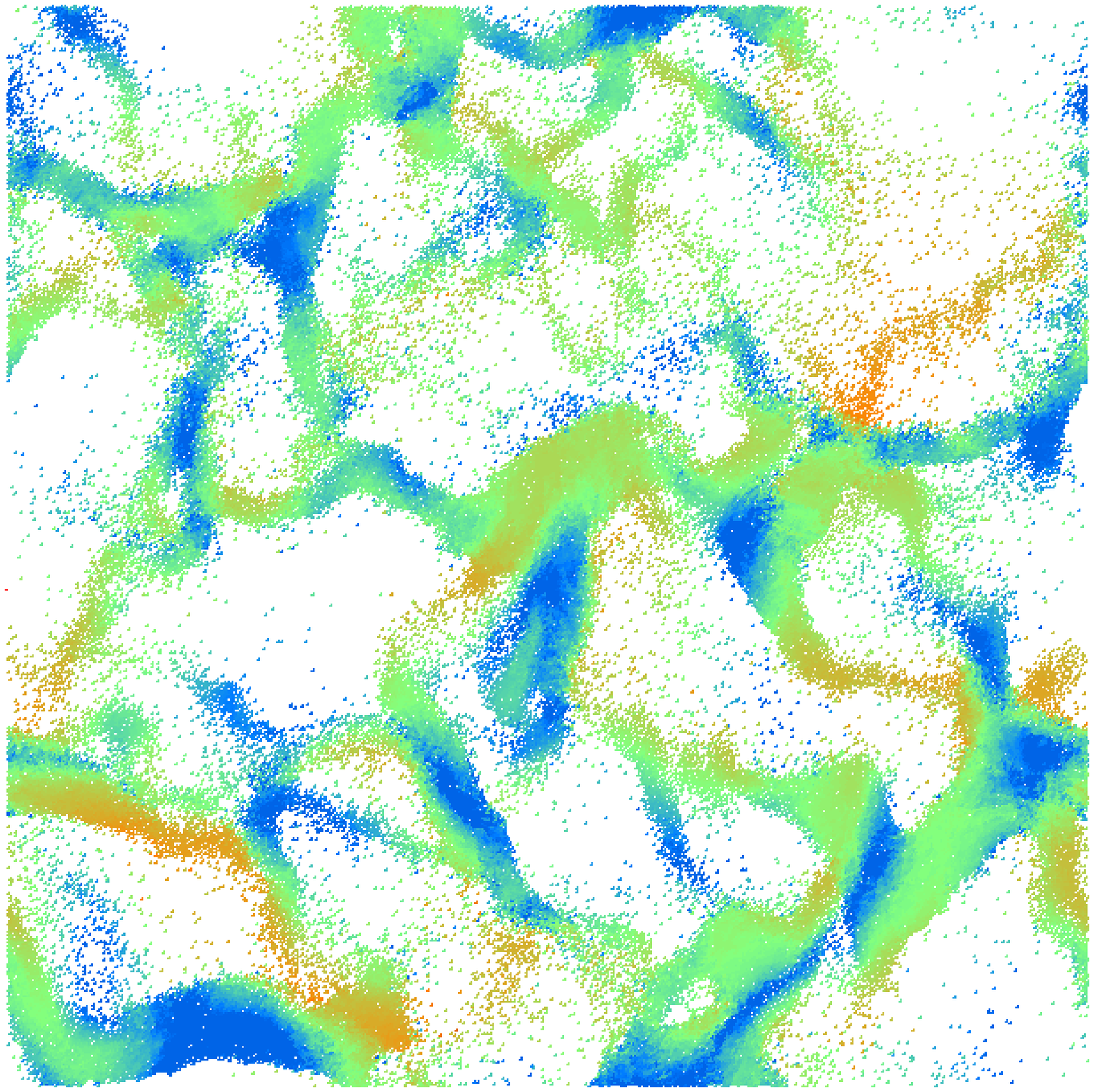,height=7.7cm} \hfill
\epsfig{file=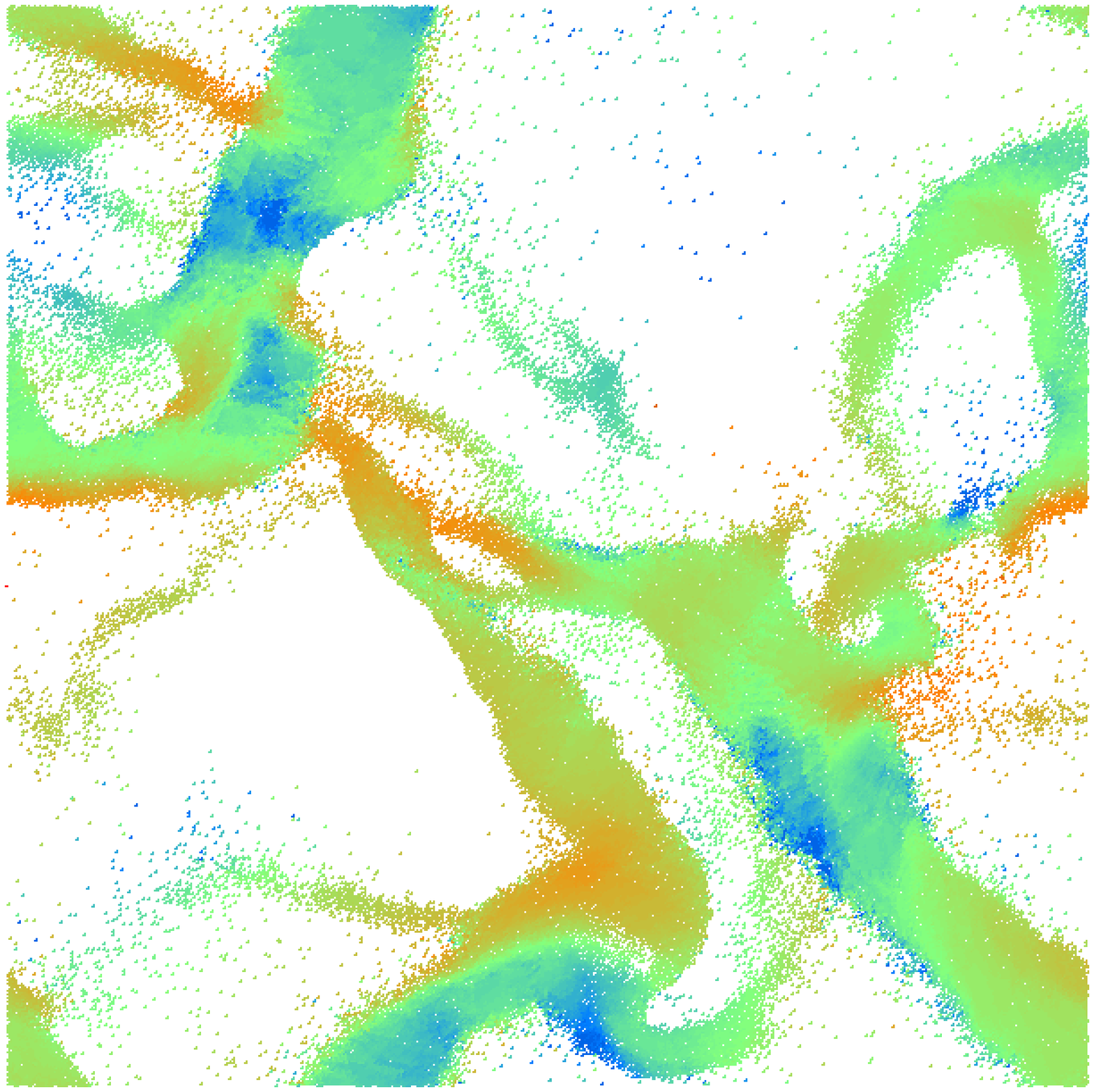,height=7.7cm}
\caption{ED simulations with $N=79524$ particles in a system of
length $L=500$ and volume fraction $\varrho = 0.25$. The restitiution
coefficient is $r=0.8$ and the critical collision frequency is
$1/t_c = 10^5$ s$^{-1}$. The energy of the single particles is 
color-coded according to the maximum (red), the mean (green), and the
minimum (blue).}
\label{fig:cluster_e}
\end{figure}

The energy $K$ decays with time, initially following the prediction for
the HCS, until $t \approx 0.2$\,s. For $t > 0.2$\,s clustering starts
and energy dissipation is slowed down. Much later, at $t \approx 50$\,s
energy decays faster again, since the clusters reach system size and 
interact via the periodic boundaries of the system. These three regimes,
which will be discussed later on, lead to the wiggly shape of the curve
$K(t)$.  The lines in Fig.\ \ref{fig:cluster_K}(a), correspond to the slopes
$-1$ and $-2$. For the homogeneous cooling state one would expect
a slope of $-2$ for long times, however, due to clustering the mean decay of 
energy is slower in the simulation. The line in Fig.\ \ref{fig:cluster_K}(b)
indicates that $C/N$ increases linearly with time, both for short and
long times. In the intermediate regime which can be identified with the
cluster growth regime the collision rate is smaller so that $C/N$ 
increases more slowly. For $t < 50$\,s, only a small percentage
of secondary collisions occur within $t_c$, but at larger times, the 
fraction of elastic collisions increases strongly.
Snapshots of the system at different times are displayed in 
Fig.\ \ref{fig:cluster_e}. 
Density variations increase with time until, at long times, the
size of the largest cluster is of order the system size.
Snapshots of the system at different times are displayed, in that 
regime, in Fig.\ \ref{fig:cluster_e}. 
The color codes
the energy in the center of mass reference frame. Thus, different
colors inside a cluster indicate either strong relative shear motion,
expansion or compression. We note that a 
cluster does not behave like a solid body, but has internal
motion and can eventually break into pieces after some time.

\begin{figure}[htbp]
{$t=40.96$ s, $C/N=670$} 
\hfill {$t=446.6$ s, $C/N=5258$} \\
\epsfig{file=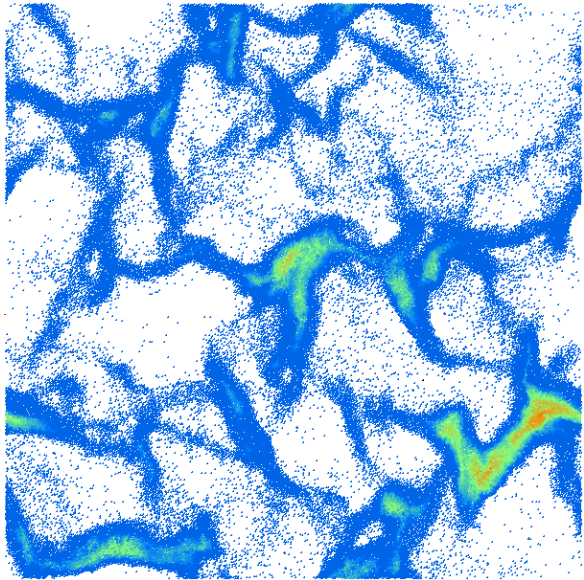,height=7.7cm} \hfill
\epsfig{file=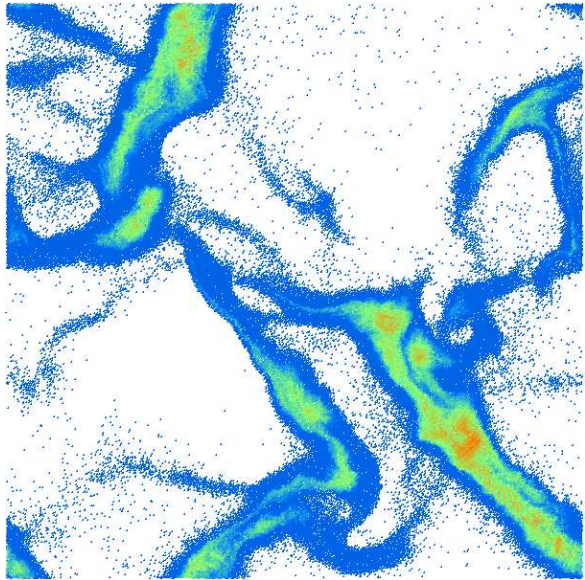,height=7.7cm}
\caption{Collision frequency of individual particles from the simulation 
in Fig.\ {\protect\ref{fig:cluster_e}}. The colors indicate large (red),
medium (green), and small (blue) collision frequencies.}
\label{fig:cluster_nc}
\end{figure}

In Fig.\ \ref{fig:cluster_nc} data from the same simulation are presented
with a different color coding. Particles with a time between collisions
smaller than $t_c$ can be seen mainly in the centers of the clusters
colored in red. Note that most of the computational effort is spent in
predicting collisions and to compute the velocities after the collisions.
Therefore, the regions with the largest collision frequencies require
the major part of the computational resources.

\begin{figure}[htbp]
{$t=40.96$ s, $C/N=670$} 
\hfill {$t=446.6$ s, $C/N=5258$} \\
\epsfig{file=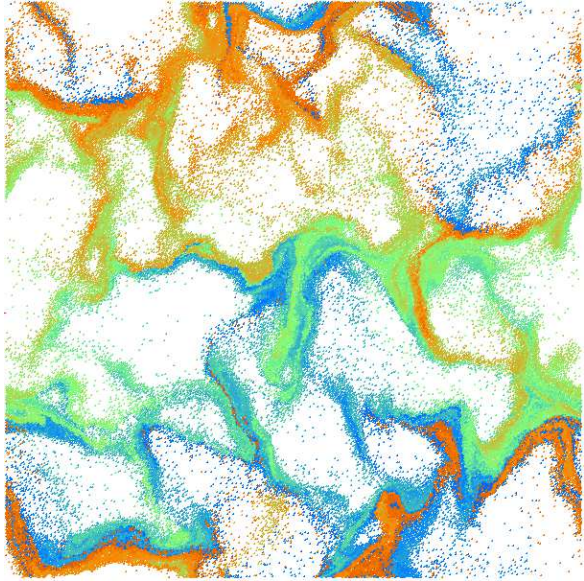,height=7.7cm} \hfill
\epsfig{file=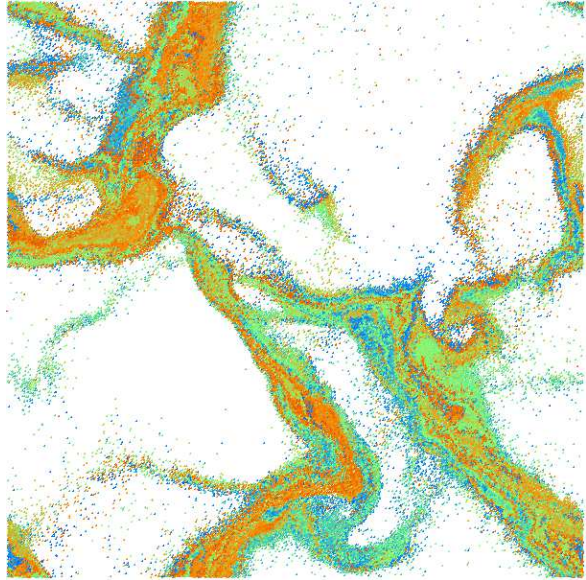,height=7.7cm}
\caption{The particles from the simulation in 
Fig.\ {\protect\ref{fig:cluster_e}} are initially colored corresponding 
to their vertical position and do not change their color. }
\label{fig:cluster_ch}
\end{figure}

An alternative color-coding of the same simulation allows us to view
the mixing in the system. Particles with similar vertical coordinates
are initially colored identically and keep their color
in Fig.\ \ref{fig:cluster_ch}. At the upper and lower boundary,
blue and red particles penetrate the others, respectively, due
to the periodic boundaries. The shape of the initially flat 
boundary-interface between red and blue particles has strong contrast
and is thus most evident.  At sufficiently large times red and blue 
particles have travelled to the center of the system, and even later, 
inside the large clusters, one observes thin stripes of one color
due to internal shear in the clusters. At the final stage, particles
coming from different origin can be very close together at any point 
in the system due to the diffusive and ballistic mixing.

\section{Quantitative description of cluster growth}
\label{sec:Burning}

In order to describe the cluster growth more quantitatively, we 
propose a method to define clusters and to identify all the 
particles belonging to a specific cluster. Similar to the so-called
``burning''-algorithm for percolation problems \cite{herrmann84b}, our method
identifies all particles that are in ``contact'' with each other. In contrast
to the traditional method, that is applied to regular lattices where
the sites are either occupied or not, our method works in the continuum
and identifies particles that are in ``contact''. Since we use
perfectly rigid particles, we have to define contact in some
arbitrary way. Two particles $i$ and $j$ with
positions $\vec r_i$ and $\vec r_j$, and diameters
$d_i$ and $d_j$, are assumed to be in contact if
\begin{equation}
|\vec r_i - \vec r_j| \le S_c (d_i+d_j)/2 ~,
\label{eq:Sc}
\end{equation}
with the distance factor $S_c > 1$.  Consequently, all third particles $k$ 
for which the condition in Eq.\ (\ref{eq:Sc}) is true relative to particle $j$, 
are also in the same cluster as particle $i$, and so on.

For the identification of clusters,
first, all particles are sorted into a ``linked-cell'' structure \cite{allen87}
what is convenient also for the simulation, and absolutely necessary
for large particle numbers $N$ to avoid the numerical effort 
{\cal O}$(N^2)$ connected to nested loops of length $N$. The size of the 
cells is larger than the maximum of all $S_c (d_i + d_j)/2$ here. From each 
cell starts a linked list that contains all particles with their 
center inside the cell. Now the criterion in Eq.\ (\ref{eq:Sc}) is 
tested for all particles $i$, but applied only to the particles $j$ 
in the same cell as $i$, or in the nearest-neighbor and 
next-nearest-neighbor cells.

The cluster identification algorithm starts with all particles $i$ 
being individual clusters of size $M_i=1$, stored in a linked list of 
length $i_{max}=N$, so that one has $N$ clusters at the beginning. 
Then, the distance between all particle 
pairs ($i$,$j$) that are close enough in the linked-cell structure is 
compared according to Eq.\ (\ref{eq:Sc}), but
only if both particles belong to different clusters $i_1$ and $i_2$.
If both particles are close enough, all particles from cluster $i_2$
are added to cluster $i_1$, while cluster $i_2$ is replaced by the last
cluster $i_{max}$. At the same time, the number of clusters $i_{max}$ is
reduced by one, so that the former cluster $i_{max}-1$ becomes the last
one in the list. This is repeated until every pair ($i$,$j$) is examined once.
If two particles belong already to the same cluster, one can proceed
with the next pair, because no cluster merging is necessary.

\begin{figure}[htb]
\begin{tabular}{ p{7.3cm} p{7.3cm} }
\epsfig{file=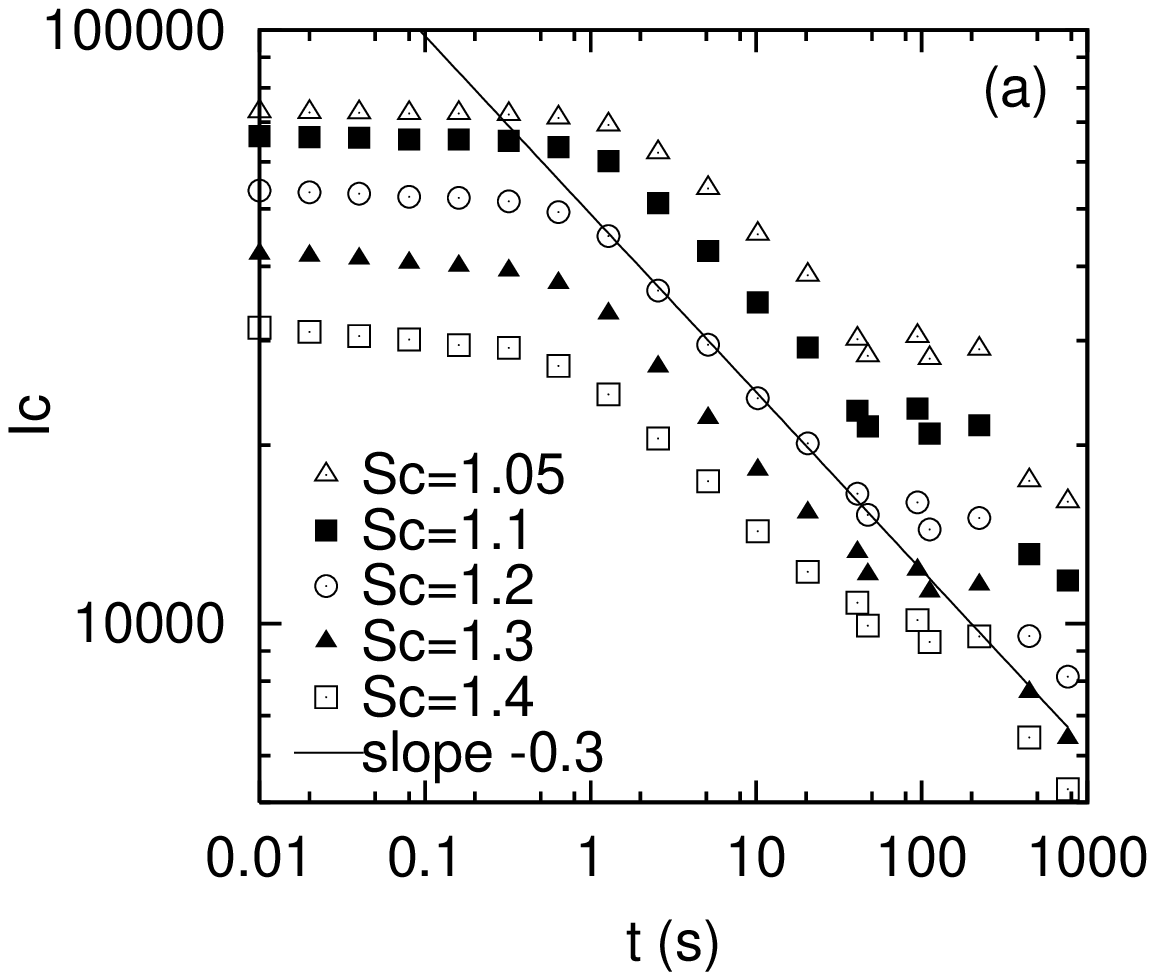,height=6.9cm,angle=-90} &
\epsfig{file=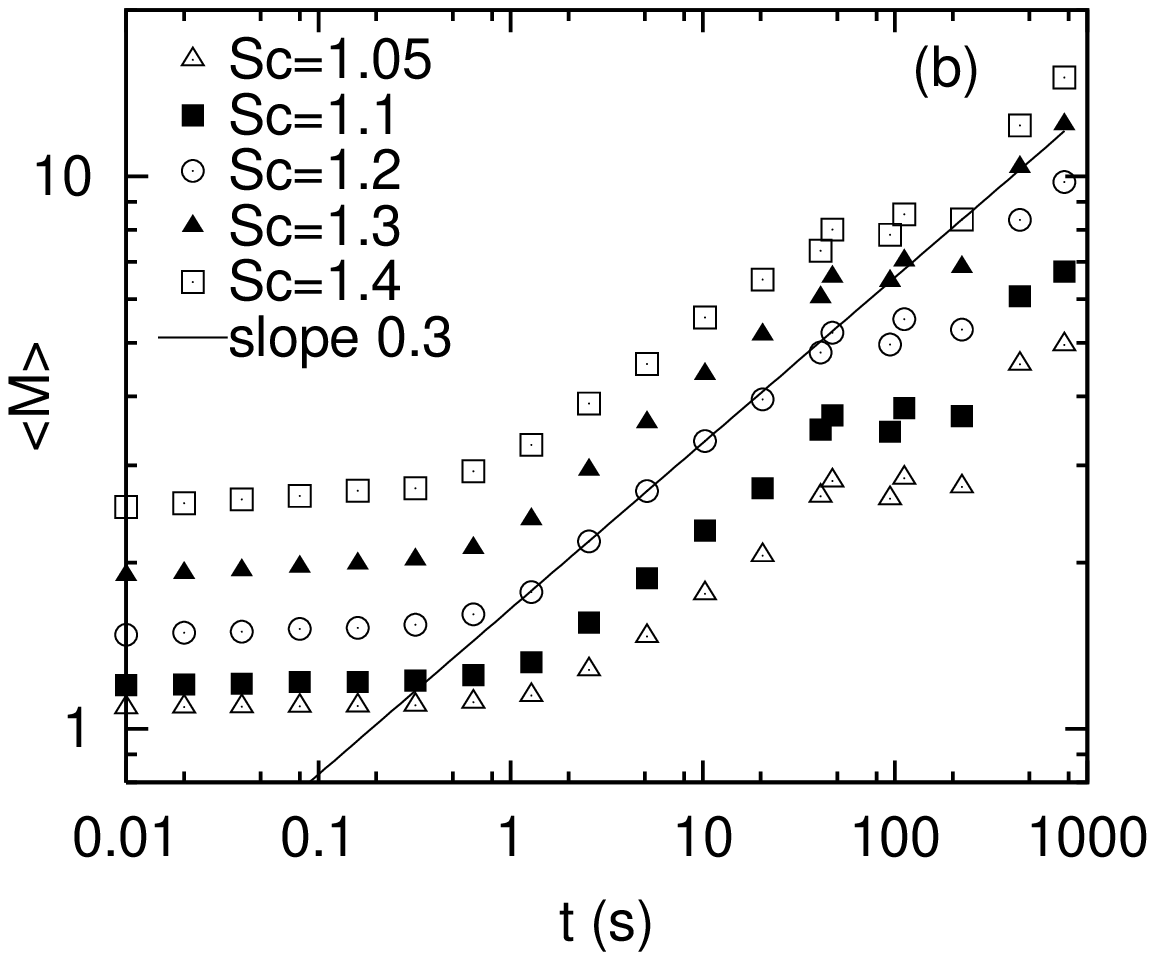,height=6.9cm,angle=-90} \\
\epsfig{file=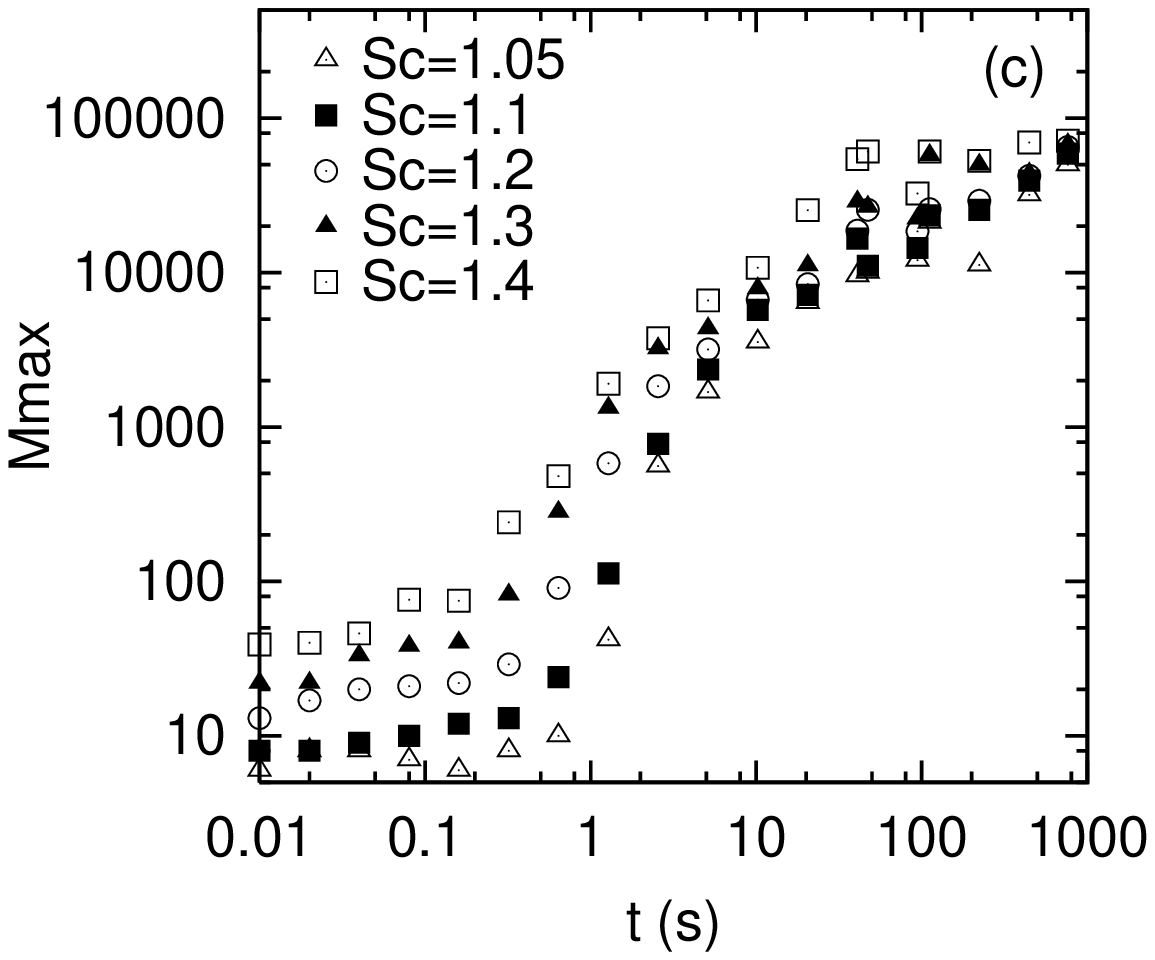,height=6.9cm,angle=-90} &
~\vspace{1.0cm}~
\caption{
(a) Number of clusters $I_c$ in the system from 
Figs.\ {\protect \ref{fig:cluster_K}-\ref{fig:cluster_ch}},
plotted against time $t$, for different $S_c = 1.05$, 1.1, 1.2, 1.3, 
and 1.4.  
(b) Mean cluster size $\langle M \rangle$ plotted against $t$.
(c) Size of the largest cluster $M_{\rm max}$ plotted against $t$.
}
\label{fig:mean}
\end{tabular}
\end{figure}

After all pairs are examined, one has  the size $M_i$ of every cluster $i$,
the number of clusters $I_c = i_{max}$, the size of the largest cluster 
$M_{max}$, and the mean cluster size
\begin{equation}
\langle M \rangle  = \frac{1}{I_c} \sum_{i=1}^{I_c} M_i ~. 
\end{equation}
The size of a cluster is in this context the number of particles
in it, and thus proportional to its mass, since we use mono-disperse 
particles. In Fig.\ \ref{fig:mean} the quantities introduced above
are displayed for the simulation in 
Figs.\ \ref{fig:cluster_K}-\ref{fig:cluster_ch}.

Since the parameter $S_c$ is arbitrary, we examine the
simulation presented above for different $S_c$ values.
The qualitative behavior is independent of $S_c$: (i) The
number of clusters decreases with time, (ii) the mean cluster
size grows correspondingly, and (iii) the largest cluster also
grows in size during the simulation.
In Figs.\ \ref{fig:mean}(a-c) it is easy to identify the three
different regimes discussed above. In the homogeneous state,
for times $t < 1$\,s, the change of $I_c$, $\langle M \rangle$, and 
$M_{\rm max}$ 
with time is very weak. At larger times, until about $t \approx 40$\,s,
the cluster growth regime is evidenced, and at very long times, one
guesses a regime where $I_c$, $\langle M \rangle$, and $M_{\rm max}$ are
again varying slowly.

Quantitatively, however, the behavior depends strongly on $S_c$.
The larger $S_c$, the higher is the probability to find a particle
$j$ that is close enough to a given particle $i$ so to be assumed 
as belonging to the same cluster. Only the maximum cluster size
at large times is a quantity that becomes rather independent of $S_c$ 
for long enough times and small enough $S_c \le 1.2$. This is because
the regions outside the largest cluster are almost empty,
so that a variation in $S_c$ has almost no effect. The few particles
that hit the largest cluster are absorbed into it with high probability,
so that it grows on and on.

The lines in Figs.\ \ref{fig:mean}(a,b) are fits of the cluster growth
behavior to the laws
\begin{equation}
I_c \propto t^{- \cal M} {\rm ~~and~} \langle M \rangle \propto t^{\cal M} ,
\end{equation}
yielding the power ${\cal M} \approx 0.3$. Thus, the behavior of $I_c$ and 
$\langle M \rangle$ is described by one power with opposite sign. 
Unfortunately, the limited system size allows a reasonable fit
only over a little more than one order of magnitude in time, so that the 
power law is a guess rather than a reliable functional behavior.

\section{Summary and Conclusion}
\label{sec:sumconc}

Simulations of freely cooling, two dimensional systems 
were presented. The behavior of the system was examined
as a function of the restitution coefficient $r$ and the 
contact duration $t_c$. The system is initially in a homogeneous
cooling state, and the deviations from this theoretically well
understood situation occur earlier with stronger dissipation.
The parameter that resembles a contact duration of the particle
pairs is $t_c$ and has -- varied over several orders of magnitude --
only a weak effect on the behavior of the system, given that it is 
essentially smaller than the typical time between collisions.

When dissipation is strong enough, density variations build up
and lead to clusters of particles. The deviation from the homogenous
cooling state goes ahead with the growth of these formations.
In order to quantify the cluster growth, we introduced a 
continuous ``burning''-algorithm that allows to access e.g.~the 
number of clusters in the system, the mean cluster size, and the 
size of the largest cluster. During the evolution of the system,
the number of clusters decreases, and the mean cluster size increases
correspondingly. The size of the maximum cluster also increases
with time, and finally most particles in the system belong to
the largest cluster.

The criterion to decide whether two particles belong to the same
cluster or not, is based on the ratio of the interparticle distance
and the particle size. Both the choice of this criterion and the
magnitude of the distance factor $S_c$ are arbitrary. Alternatives
would involve the collision frequency or the relative velocity of
the particles under examination. However, we prefer the distance criterion
for several reasons: (i) If $S_c$ is neither too small nor too large
($1.01 \le S_c \le 1.2$ leads to reasonable results), the criterion
accounts for nearest neighbors in a close packing, and particles that 
``feel'' each other in the sense that 
a shear displacement would need a dilation of neighboring layers.
(ii) The criterion is independent of the dimension (it is easily applied
also to 3D situations -- data not shown here), and (iii) the
distance criterion can be applied to snapshots from a simulation,
whereas other criteria might need more data and information. 
(iv) Finally, the distance
criterion is independent of initial conditions like temperature, in 
contrast to a criterion that is based on the relative velocities of 
particles or on their collision frequencies.

We conclude that even about $10^5$ particles are not enough to examine the
cluster growth regime over long enough times. Possibly, much larger simulations
are necessary, in order to extend the available range in time. The presented
large simulation was carried out on a small IBM 43P Power PC with 64 MB memory 
in several days of CPU time, so that a factor of at least 10 in the particle 
number should be possible with a faster computer. The event driven algorithm 
used is optimized for scalar machines, so that parallelization or 
vectorization are not feasible in the present state. 
However, there exist alternative 
schemes like the DSMC simulation method \cite{muller97,luding98b,luding98e},
which can be parallelized so that much larger particle numbers are
accessible.

On the other hand, a more detailed parameter study, involving different 
collision models and possibly rotational degrees of freedom, might uncover
parameter combinations with much slower cluster growth. The few simulations, we
performed with different restitution coefficients and also with more realistic
collision laws that couple to the rotational degrees of freedom 
\cite{luding98c}, however, did not lead to remarkably different behavior.
A more detailed parameter study and also the analysis of the cluster-size
probability distribution in 2D and 3D are future perspectives requiring the
use of a cluster-identification algorithm as proposed in this study.

\section*{Acknowledgements}

We acknowledge the support of the the Deutsche Forschungsgemeinschaft (DFG), 
Sonderforschungsbereiche (SFB) 381 and 382.


\end{document}